\documentclass[10pt,conference]{IEEEtran}
\IEEEoverridecommandlockouts
\usepackage{cite}
\usepackage{amsmath,amssymb,amsfonts}
\usepackage{algorithmic}
\usepackage{graphicx}
\usepackage{textcomp}
\usepackage{xcolor}
\usepackage{tabularx}
\usepackage{booktabs}       
\def\BibTeX{{\rm B\kern-.05em{\sc i\kern-.025em b}\kern-.08em
    T\kern-.1667em\lower.7ex\hbox{E}\kern-.125emX}}
\begin{document}

\title{
Retrieve and Refine: Exemplar-based Neural Comment Generation
}
\author{\IEEEauthorblockN{Bolin Wei}
		\IEEEauthorblockA{\textit{Key Laboratory of High Confidence Software Technologies (Peking University)}\\
    \textit{Ministry of Education, China; Software Institute, Peking University, China}\\
		bolin.wbl@gmail.com}
}

\maketitle

\begin{abstract}
Code comment generation is a crucial task in the field of automatic software development. Most previous neural comment generation systems used an encoder-decoder neural network and encoded only information from source code as input. Software reuse is common in software development. However, this feature has not been introduced to existing systems. Inspired by the traditional IR-based approaches, we propose to use the existing comments of similar source code as exemplars to guide the comment generation process. Based on an open source search engine, we first retrieve a similar code and treat its comment as an exemplar. Then we applied a seq2seq neural network to conduct an exemplar-based comment generation. We evaluate our approach on a large-scale Java corpus, and experimental results demonstrate that our model significantly outperforms the state-of-the-art methods.
\end{abstract}


\section{Introduction}
Code comments provide a clear description for a piece of the source code, which can help programmers understand programs quickly and correctly~\cite{DBLP:conf/kbse/SridharaHMPV10} and are beneficial for software maintenance~\cite{DBLP:conf/sigdoc/SouzaAO05}. However, generating proper comments is time-consuming. Hence, carrying out automatic comment generation becomes greatly crucial for software development.

Recently, many researchers have applied encoder-decoder neural networks and achieved state-of-the-art performance on the comment generation~\cite{DBLP:conf/iwpc/HuLXLJ18, DBLP:conf/kbse/WanZYXY0Y18}. It's worth noting that their methods only take the source code information as input. However,
we find that these models tend to generate short comments and produce degenerate outputs in a state of ``out of control". 
For example, on our Java dataset,
about 2\% of predictions by DeepCom~\cite{DBLP:conf/iwpc/HuLXLJ18} are less than four words, and there are over seventy HAD's~\cite{DBLP:conf/kbse/WanZYXY0Y18} predictions repeating a token for 
twenty times. Therefore, we argue that it is not enough to generate comments only based on the source code.

\begin{figure}[!t]
	\centering
	\resizebox{.7\linewidth}{2.5cm}{
	\includegraphics[width=\linewidth]{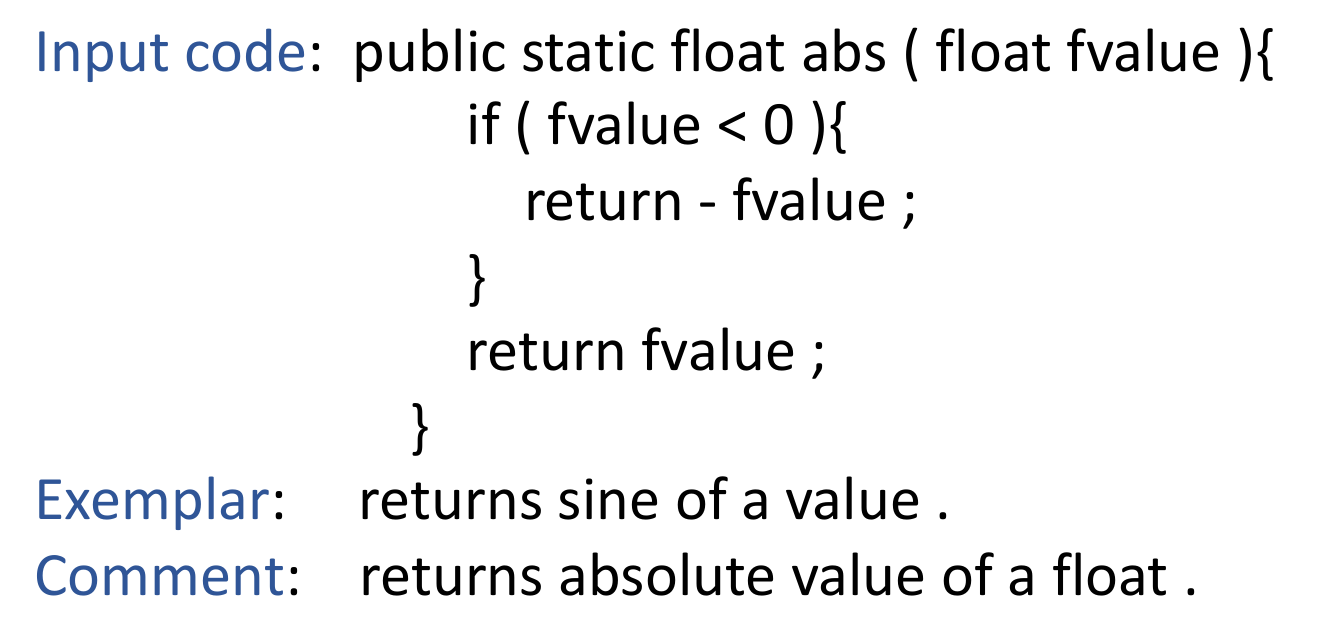}
	}\vspace{-.2cm}
	\caption{A code-comment pair from our dataset, along with its exemplar.}\label{code}\vspace{-.4cm}
\end{figure}

Software reuse is common in software development~\cite{DBLP:conf/msr/WangDZCXZ13, DBLP:conf/ecoop/ZhongXZPM09}, that is, for a given input code, we can search for a similar piece of code in a large-scale software library. If source code in this library contains comments, then we can use these comments to assist in the comment generation for the given code. The traditional IR-based comment generation are based on this idea~\cite{DBLP:conf/icse/HaiducAM10, DBLP:conf/wcre/HaiducAMM10, DBLP:conf/wcre/WongLT15, DBLP:conf/kbse/WongYT13} and treat the retrieved comments directly as comments for a given code. However, 
the retrieved comment may contain some information that is inconsistent with the content in the given code. A better way is to treat comments of similar code as an exemplar to guide the comment generation. Figure~\ref{code} shows a simple code-comment pair together with its exemplar from our dataset. 

Exemplar-based comment generation can be seen as a template-based comment generation process. 
However, defining a template is a time-consuming task and requires some domain knowledge. By using comments of similar code as exemplars, on the one hand, the workload of constructing templates is reduced, and on the other hand, the process of comment generation is controlled.

In this paper, we present an exemplar-based comment generation model, which consists of two modules, Retrieve and Refine. In Retrieve, we exploit an open source search engine to retrieve the similar code. In Refine, we use LSTMs~\cite{DBLP:journals/neco/HochreiterS97} to encode the given code, the similar code and the comment of the similar code into feature vectors. Based on the correlation of the given code and similar code, a new comment is generated concerning the comment of the similar code.

\section{Background and related work}
The automatic comment generation is a popular research task in the field of software engineering. Researchers have proposed various approaches to solve this task, including manual templates, IR, and deep learning-based methods. Template-based approaches~\cite{DBLP:conf/iwpc/MorenoASMPV13, DBLP:conf/kbse/SridharaHMPV10, DBLP:conf/icse/SridharaPV11, DBLP:conf/iwpc/McBurneyM14} generate comments based on manually defined rules and information from the source code. IR-based methods~\cite{DBLP:conf/wcre/HaiducAMM10, DBLP:conf/iwpc/EddyRKC13, DBLP:conf/wcre/WongLT15, DBLP:conf/kbse/WongYT13} include retrieving terms using text retrieval techniques or retrieving similar code using code clone detection techniques, then combining terms into comments or treating comments of similar code directly as comments. Deep learning-based approaches~\cite{DBLP:conf/iwpc/HuLXLJ18, DBLP:conf/kbse/WanZYXY0Y18, DBLP:conf/icml/AllamanisPS16, DBLP:conf/acl/IyerKCZ16} are data-driven approaches that rely on large-scale code-comment pairs. The existing methods mainly utilize an encoder-decoder neural network, the encoder models the source code sequence or the AST of source code, and the decoder generates comments. However, the existing deep learning-based models do not take advantage of the feature of software reuse and lack control when generating comments, making it easy to produce short and meaningless comments.

\begin{figure}[t]
	\centering
	\resizebox{.9\linewidth}{!}{
	\includegraphics[width=\linewidth]{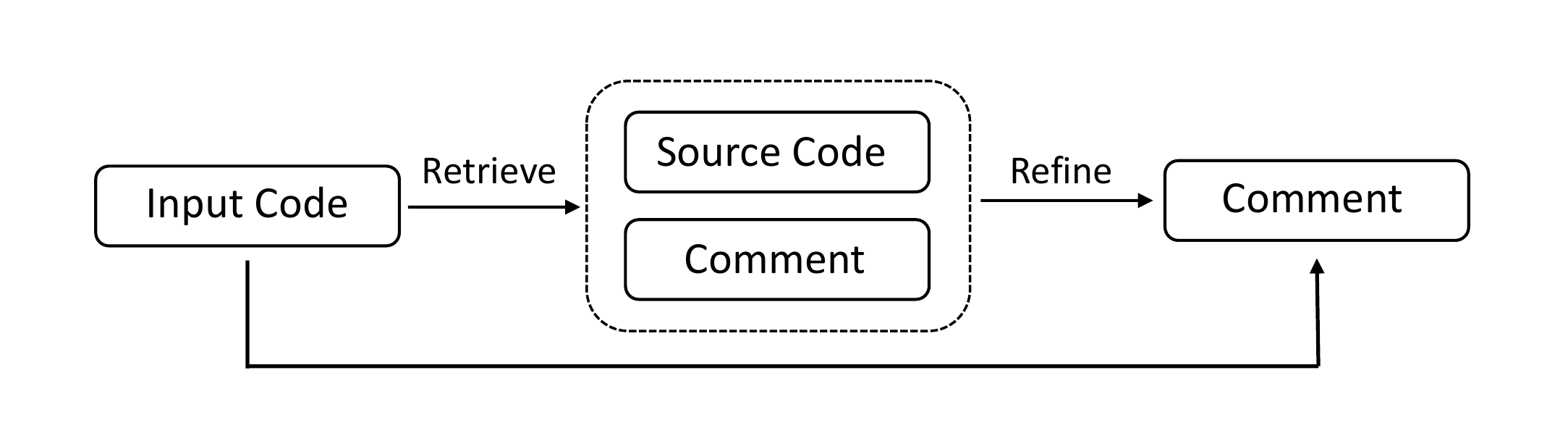}
	}\vspace{-.2cm}
	\caption{An overview of our approach for exemplar-based comment generation. We use a dashed box to represent the pair of similar code and comment.}\label{fig:framework}\vspace{-.2cm}
	\label{framework}
\end{figure}

\section{Approach overview}
Our approach is shown in Figure~\ref{framework}, which consists of two parts, a Retrieve module for searching for similar code and a Refine module for comment generation.

\subsection{Retrieve Module}
The purpose of the Retrieve module is to find an appropriate exemplar in the training set. We assume that comments of similar code have similar patterns. Thus the process of finding a proper exemplar can be transformed into the process of finding a similar code. Specifically, given a piece of the input code, we search for code similar to it and treat a comment of the similar code as an exemplar. 
In this study, we leverage an open source information retrieval engine Lucene\footnote{https://lucene.apache.org/} to build our code library and keep the default settings\footnote{TextField with WhitespaceAnalyzer} in Lucene. For each piece of code, we chose the first-ranked similar code and used its comment as an exemplar.

\subsection{Refine Module}
Once we have an exemplar, a straightforward way is to use it as a comment for the input code. However, the code usually contains information that is not in the input code, such as different variables. Therefore, a neural network can be applied to learn the difference between the input code and the similar code to improve the exemplar.

Specifically, the Refine module utilizes a seq2seq neural network~\cite{DBLP:conf/nips/SutskeverVL14}. The encoder uses three LSTMs to encode the input code, the similar code, and the comment of the similar code and explores the difference between the source code and the similar code using a nonlinear \textit{sigmoid} function to obtain a similarity score $s$. The decoder, another LSTM with the attention mechanism~\cite{DBLP:journals/corr/BahdanauCB14}, generates a new comment whose initial hidden state is the combination of the last hidden states of the input code $\textbf{h}^x$ and the comment of the similar code $\textbf{h}^y$:
\begin{align*}
    \textbf{h}^x*(1-s)+\textbf{h}^y*s
\end{align*}
The purpose of this combination is that if the input code is very different from the similar code, that is, the similarity score is lower, then the decoder should pay more attention to the content of the input code. For each output token, the attention mechanism calculates its similarity to all input code tokens and its similarity to all comment tokens of the similar code and obtains two context vectors. Then we use $s$ to sum the two context vectors and get the final context vector.

During inference, we use a beam search～\cite{DBLP:conf/nips/SutskeverVL14} to generate comments. The training objective is the cross-entropy between the prediction and the ground truth.

\begin{table}[t]
\small
\caption{Evaluation results on Java methods}
\centering
\begin{tabular}{lcc}
\toprule
{\bf Methods} & BLEU (\%) & METEOR (\%) \\ \midrule
Vanilla seq2seq   & 30.90   &   18.85     \\
DeepCom   & 32.30   & 20.74   \\
HAD   &  33.24  &  21.27   \\
Our model &  {\bf 43.78}  &  {\bf 28.12}   \\ \bottomrule
\end{tabular}
\label{results}
\vspace{-.2cm}
\end{table}

\section{experiments}
We collected a large number of Java projects from GitHub, extracted Java methods from them, and used the first sentence of Javadoc as a comment for the method. We processed the dataset as in ~\cite{DBLP:conf/iwpc/HuLXLJ18}. After eliminating the duplicate data samples, we can obtain about 370k $\langle \text{method}, \text{comment} \rangle$ pairs. The training, validation, and test sets are about 330k, 20k, and 20k, respectively. We evaluate the performance based on BLEU~\cite{DBLP:conf/acl/PapineniRWZ02} and METEOR~\cite{DBLP:conf/acl/BanerjeeL05}, which are used in ~\cite{DBLP:conf/iwpc/HuLXLJ18, DBLP:conf/kbse/WanZYXY0Y18}.

We compared our approach to three neural network-based baseline methods, including vanilla seq2seq~\cite{DBLP:conf/nips/SutskeverVL14}, DeepCom~\cite{DBLP:conf/iwpc/HuLXLJ18}, and HAD~\cite{DBLP:conf/kbse/WanZYXY0Y18}. The last two methods are state-of-the-art methods on comment generation. Table~\ref{results} illustrates the BLEU and METEOR scores of different methods. 
As can be seen from the results, our approach is significantly better than all baseline methods on both metrics. The BLEU score of our model improves about
10\% compared to HAD. Without similar code and comments, our model is equivalent to the vanilla seq2seq;
hence, it is convincing that exemplars are very helpful for generating comments.

\section{conclusion and future work}
This paper creatively proposes a method for automatically generating comments based on exemplars, which includes two modules. The Retrieve module finds out a comment of a similar code as an exemplar. The Refine module exploits the correlation of the input code and the similar code to generate comments under the guidance of the exemplar. Experimental results show that our method can significantly outperform all previous state-of-the-art methods on the Java dataset.

This work has some limitations. On the one hand, our Retrieve module leverages an open source search engine. While ensuring efficiency, the token-based code search technique is too simple to retrieve the semantically similar code. In the future, we will use advanced code clone detection to improve the performance of the Retrieve module. 
On the other hand, the evaluation of the prediction is now based on metrics in machine translation. We expect to use human evaluation approaches to measure whether our methods can be applied in the software development process.

\bibliographystyle{IEEEtran}
\bibliography{ref}

\end{document}